\documentclass[aps,prd,floatfix,twocolumn,superscriptaddress,preprintnumbers,nofootinbib,10pt]{revtex4}
\usepackage{slashed}

\usepackage{epsfig,latexsym,cancel,amssymb,amsmath,verbatim,mathrsfs}
\usepackage{color}
\usepackage{graphicx}

\def\ra{\rightarrow}
\def\L{\left(}
\def\R{\right)}
\def\wt{\widetilde}

\def\ld{\lambda}
\def\f{\frac}
\newcommand{\be}{\begin{equation}}
\newcommand{\ee}{\end{equation}}
\newcommand{\bea}{\begin{eqnarray}}
\newcommand{\eea}{\end{eqnarray}}
\newcommand{\ba}{\begin{array}}
\newcommand{\ea}{\end{array}}

\long\def\symbolfootnote[#1]#2{\begingroup%
\def\thefootnote{\fnsymbol{footnote}}\footnote[#1]{#2}\endgroup}

\newcommand{\beq}{\begin{equation}}
\newcommand{\eeq}{\end{equation}}

\begin{document}

\title{A Rediatively Light Stop Saves the Best Global Fit for Higgs Boson Mass and Decays}

\author{Zhaofeng Kang}

\affiliation{State Key Laboratory of Theoretical Physics
and Kavli Institute for Theoretical Physics China (KITPC),
Institute of Theoretical Physics, Chinese Academy of Sciences,
Beijing 100190, P. R. China}
\affiliation{Center for High-Energy Physics, Peking University, Beijing, 100871, P. R. China}

\author{Tianjun Li}

\affiliation{State Key Laboratory of Theoretical Physics
and Kavli Institute for Theoretical Physics China (KITPC),
Institute of Theoretical Physics, Chinese Academy of Sciences,
Beijing 100190, P. R. China}

\affiliation{George P. and Cynthia W. Mitchell Institute for
Fundamental Physics and Astronomy, Texas A$\&$M University,
College Station, TX 77843, USA}

\author{Jinmian Li}

\affiliation{State Key Laboratory of Theoretical Physics
and Kavli Institute for Theoretical Physics China (KITPC),
Institute of Theoretical Physics, Chinese Academy of Sciences,
Beijing 100190, P. R. China}

\author{Yandong Liu}

\affiliation{State Key Laboratory of Theoretical Physics
and Kavli Institute for Theoretical Physics China (KITPC),
Institute of Theoretical Physics, Chinese Academy of Sciences,
Beijing 100190, P. R. China}

\date{\today}

\begin{abstract}

The LHC discovered the Standard Model (SM) like Higgs boson with mass
around 125~GeV. However, there exist hints of deviations from Higgs
decays. Including the Tevatron data, the deviations can be
explained by the extremely mixed stop sector in the sense of best global fit (BGF).
We analyze the relations among the competing reduced coupling $hGG$, Higgs boson mass,
and LHC stop mass $m_{\wt t_1}$ lower bound at the tree- and one-loop level.
In particular, we point out that we use the light stop running mass in the Higgs
boson mass calculation while the light stop pole mass in the Higgs decays. So
the gluino radiative correction on the light stop mass plays the crucial role.
Its large negative correction saves the GBF in the Minimal Supersymmetric SM (MSSM) and the
next to the MSSM (NMSSM) constrained by
the perturbativity. Moreover, a light stop is predicted: in the MSSM if we set the gluino mass
$M_3\lesssim4$~TeV, we have $m_{\wt t_1}<m_t$; while in the NMSSM, imposing the least tuning
condition we get $m_{\wt t_1}\simeq130$~GeV for $M_3\simeq1.5$~TeV.

\end{abstract}

\pacs{}
\maketitle

\noindent {\bf{Introduction:}}
The LHC has not found any direct signature of supersymmetry (SUSY) so far.
But at the 5$\sigma$ level the discovery of the Standard Model (SM)-like Higgs
boson with mass $m_h\simeq$125~GeV~\cite{:2012gk, :2012gu},
indirectly supports SUSY from several aspects.
In the first, if this spin-0 particle is confirmed as a fundamental
particle,  we must explain the corresponding quardratic divergency problem,
and SUSY is the most elegant solution. Next, one may wonder why the Higgs
boson is so light, {\it i.e.}, why is it not 150~GeV or even heavier? But this is not a
puzzle in the SUSY models, which actually are suffering from the opposite puzzle:
why is the Higgs boson so heavy?

Last but never the least, the discovered Higgs boson is not exactly SM-like
and shows some notable deviations which may imply new physics.
Both the gluon fusion (GF) and vector boson fusion (VBF) production modes
show significant di-photon excess.
For the $pp\ra h\ra V^*V$ channel there exists a mild
deficit in the CMS experiment~\cite{:2012gu} but
small excess in the ATLAS experiment~\cite{:2012gk}.
Including the Tevatron data, the best global fit (BGF) gives~\cite{Buckley:2012em}
 {\small\begin{align}\label{cg}
&c_{g}^2=\f{\Gamma(h\ra gg)}{\Gamma^{\rm SM}(h\ra gg)}\approx0.7, \,\,
 c^2_{\gamma}=\f{\Gamma(h\ra \gamma\gamma)}{\Gamma^{\rm SM}(h\ra \gamma)}\approx 2.1,
  \end{align}}
with all the other couplings fixed. The reduced couplings $c_{g,\gamma}$
   are sensitive to colored and/or charged particles beyond the SM,
and thus indicating new physics.

In SUSY models, stau can increase $c_\gamma$ but does not affect $c_g$~\cite{stau}
while the stops, which carry the $SU(3)_C$ and $U(1)_{EM}$ quantum numbers  simultaneously,
may alter the reduced couplings towards Eq.~(\ref{cg}).
Practically, if the stop-loop interferes destructively with the top-loop
contributing to $h\ra gg$ while constructively with the $W-$loop
contributing to $h\ra \gamma\gamma$, then the stops can improve
the global fit by an amount $\Delta\chi^2\sim10$ (compared to the SM fit).
However, the above BGF in Eq.~(\ref{cg}) can not be realized generically in the SUSY models
due to the relatively large SM-like Higgs boson mass.
In this letter, we investigate the
possibility of accommodating such BGF in the conventional
minimal supersymmetric SM (MSSM) and the next-to-the MSSM (NMSSM)~\cite{Ellwanger:2009dp}.
We find that the BGF can be obtained if the gluino radiative correction
to the light stop pole mass is negative and large.

\noindent {\bf{$c_{g}$ competing with $m_h$ and $m_{\wt t_1}$:}} The stop sector is the
focus of SUSY phenomenology right now due to various reasons.
Firstly, the stop corrections~\cite{Higgs:MSSM} lift $m_h$ above the LEP bound and
even to 125~GeV indicated by the LHC. At the one-loop level, the corrections can be
approximated as
 \begin{align}\label{1loop}
\Delta_t\simeq\f{3m_t^4}{4\pi^2 v^2}\L \log\L\f{m_{\wt t}^2}{m_t^2}\R+\eta_t^2\L1-\f{\eta_t^2}{12}\R\R,
  \end{align}
with $v=174.6$~GeV, $\eta_t=X_t/m_{\wt t}$ with $X_t\equiv A_t-\mu\cot\beta$
the stop mixing parameter, and $m_{\wt t}=\sqrt{m_{\wt t_1}m_{\wt t_2}}$ the geometric mean of
two stop masses. Here, $A_t$ is the top quark trilinear soft term and $\mu$ is
the Higgs bilinear mass term in superpotential.
When $\eta_t\ll 1$, the mixing term is ignorable and
the leading logarithm dominates the corrections. While in the maximal
mixing scenario with $\eta_t\approx \sqrt{6}$~\cite{Brummer:2012ns}, the mixing
term effect maximizes. However, as $\eta_t>2\sqrt{3}$ this effect begins to
act reversely. We will find that it is the typical case
to approach the BGF.

Next, the stop sector can change the reduced couplings of $hGG$ and $h\gamma\gamma$
simultaneously. In terms of the BGF in Eq.~(\ref{cg}), the stop-loop should destructively
interfere with the top-loop and thus modify the total reduced $hGG$ coupling to be~\cite{Blum:2012ii}
 \begin{align}\label{rGG}
c_g\approx 1+\f{1}{4}\L\f{m_t^2}{m_{\wt t_1}^2}-\eta_t^2 \f{m_{t}^2}{m_{\wt t}^2}\R\sim-0.84,
  \end{align}
where the heavier stop contribution is ignored. Accordingly,
the reduced $h\gamma\gamma$ coupling is modified to be~\cite{Blum:2012ii}
 \begin{align}\label{rgamma}
c_\gamma\approx 1.28-0.28c_g~,
  \end{align}
which is valid for $m_h\simeq125$~GeV and $c_V\simeq1$.

Finally, the stops carry colour charges and have a close relation
with the pattern of electroweak symmetry breaking. Thus, they are among the main
search particles at the LHC. Presently, the lighter stop is bounded by
$m_{\wt t_1}\gtrsim130$~GeV  (details is postponed).
Then we get $m_{\wt t_L}\gtrsim600$~GeV
from the approximation which is valid for highly mixed stops:
      \begin{align}\label{lowerbound}
m_{\wt t_L}\approx (2C_G)^{1/2}m_{\wt t_1},\quad |X_t|\approx
(2C_G)^{1/2}\f{m_{\wt t_1}}{m_t}m_{\wt t_L},
  \end{align}
with $C_G\approx4(1-c_g)+m_t^2/m_{\wt t_1}^2$. We have taken the
simplifying assumption $m_{\wt t_R}=m_{\wt t_L}$, but it can be abandoned.

Synthesizing these aspects, we obtain the competing relations among $c_g$ and $m_h$,
$m_{\wt t_1}$, which make it difficult to realize the BGF in the (N)MSSM.
(A) At least in the MSSM, to get $m_h\simeq125$~GeV a large $m_{\wt t}$ and properly
mixed stop sector are necessary. It means that generically the mixing term in
Eq.~(\ref{rGG}) is ignorable, while the lighter stop enhances $c_g$ instead of
the other way around. (B) The NMSSM, in spite of its advantage in lifting $m_h$ which
weakens the relation with the stop sector, is also impossible. This is blamed to
the lower bound on $m_{\wt t_L}$, which
leads to a big suppressing $m_t^2/m_{\wt t}^2\ll 1$ in $c_g$. Consequently, if
the mixing term can still make $c_g\sim-0.8$, we likely need $\eta_t>12$ to
compensate for that. In turn, such a large mixing term, which grows quarticly
in $\Delta_t$, will decrease $m_h$. However,
 this behavior is only an approximation, and it underestimates $\Delta_h$
in the large mixing region. In the numerical plot we will adopt the complete expression.

\noindent {\bf{A radiatively light stop mass as a savior:}}  The previous robust relations
from the two-parameters stop system can be relaxed, if we take into account the sizeable
one-loop radiative correction to the lighter stop (the much smaller correction to the heavier
stop is neglected here). Such correction is crucial, since the stop mass used to calculate
$\Delta_t$ is the running $\overline{\rm DR}$ mass, while the stop mass in Eq.~(\ref{rGG}),
used to calculate $c_g$, is the pole mass. They are related by the equation
      \begin{align}\label{}
m_{\wt t_1}^2=m_{\wt t_1}^2(Q)-{\rm Re}(\Pi(p^2,Q))|_{p^2=m_{\wt t_1}^2},
  \end{align}
where the running mass $m_{\wt t}(Q)$ is defined at the renormalization scale
$Q=m_{\wt t_L}$. $\Pi(p^2,Q)$ is the self-energy, dominated by the QCD and top-Yukawa
radiative corrections from the gluon, gluino, stop and chargino loop, etc. When the
positive self-energy is sufficiently large, the pole mass can be much lighter than the running mass.

The gluino mass plays the primary role to reduce the running mass. There exists
some uncertainties from the Yukawa and two-loop corrections. In the analytical analysis
we will not go through the details of this discussion, but instead introduce an extra
parameter $\omega_t$, the ratio of pole mass and running mass, to account for those.
Illustratively, at the order ${\cal O}(\alpha_s)$, $\omega_t$ is roughly given by~\cite{Pierce:1996zz}
\begin{align}\label{}
\omega_t\approx1 + &\frac{2}{3} \frac{\alpha_{s}}{\pi} \left[ \frac{2M_3^2}{m_{\tilde{t}_{1}}^2(Q)} \L1- \ln\frac{M_3^2}{Q^2}\R
 \right.& \nonumber \\
 &\left.
 - \frac{m_{\tilde{t}_{2}}^2(Q)}{2m_{\tilde{t}_{1}}^2} \L1
 -\ln\frac{m_{\tilde{t}_{2}}^2(Q)}{Q^2}\R
\right]_{Q=m_{\tilde{t}_{L}}}
\end{align}
The approximation is valid for $M_3,m_{\wt t_2}(Q)\gg m_{\wt t_1}(Q)\sim m_t$. Note that we have
$m_{\wt t_2}>Q$. Given a heavy gluino (required by the LHC bound), the running mass can
readily be reduced by a factor 2 or 3. As a consequence, one can work in the maximal mixing
scenario $\eta_t^2(m_{\wt t_L})\simeq6$, while $\eta_t^2(\rm pole)$ is several times larger
than $\eta_t^2(m_{\wt t_L})$. In this way, the radiatively reduced stop mass offers a solution
to reconcile the serious tension between $m_h$ and $c_g$.
 \begin{figure}[htb]
\begin{center}
\includegraphics[width=2.5in]{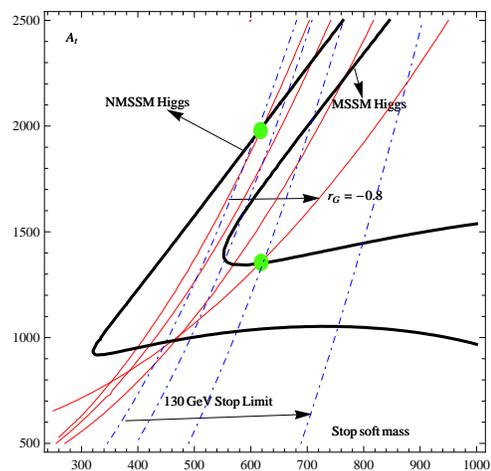}
\end{center}
\caption{\label{rGmh} Schematic plots of $m_h=125$~GeV, $c_g=-0.8$ (red thin solid lines)
and $m_{\wt t_1}=130$~GeV (dashed lines) on the $m_{\wt t_L}-X_t$ plane.
For comparison, we choose $\omega_t=0.5,0.4,0.3,0.2$ for $c_g$ and $m_{\wt t_1}$, respectively
from the left to right. In plotting $m_h$ we have fixed $\tan\beta=25$ in the MSSM
and $\tan\beta=2,\,\ld=0.6$ in the NMSSM. The running top mass is taken to be 155~GeV.}
\end{figure}

The schematic analysis in Fig.~\ref{rGmh} manifests the point. In the MSSM, as $\omega_t$
decreases the $c_g=-0.8$ curves, which are denoted by the red solid lines, move towards
the left and meet with the $m_h=125$~GeV curve at a large $m_{\wt t_L}$. As for the NMSSM,
such meeting still needs $w_t\lesssim0.6$ (we will turn back to this point later).
In this case, the lower bound on the stop pole mass $m_{\wt t_1}=$130~GeV means
that the running mass is above 220~GeV. In turn, we require a rather heavy gluino mass to
account for the big gap. Additionally, from Fig.~\ref{rGmh} one can see that for a given
$\omega_t$, the $c_g=-0.8$ curve intersects with the $m_{\wt t_1}=$130~GeV curve at large
$m_{\wt t_L}$ (understood similarly to Eq.~(\ref{lowerbound})). And its value increases as $w_t$ decreases.

$w_t\lesssim0.6$ is owing to the NMSSM
perturbativity up to $M_{\rm GUT}$. It renders $\ld\lesssim0.7$ and
imposes the tree-level bound $m_h\lesssim\ld v\sin\beta\simeq110$GeV.
Then a moderate stop correction and/or pushing effect is still indispensable.
In turn, in light of the relations between competing $m_{\wt t_1}$, $m_h$ and $c_g$,
the NMSSM with $w_t\sim1$ fails. But in $\ld$SUSY~\cite{ldSUSY}
with $\ld\sim2$, or other non-decoupling $F/D-$term models~\cite{Cheung:2012zq},
tree-level $m_h$ can be sufficiently heavy such that large $\eta_t(>12)$ is endured.

 \begin{figure}[htb]
\begin{center}
\includegraphics[width=2.6in]{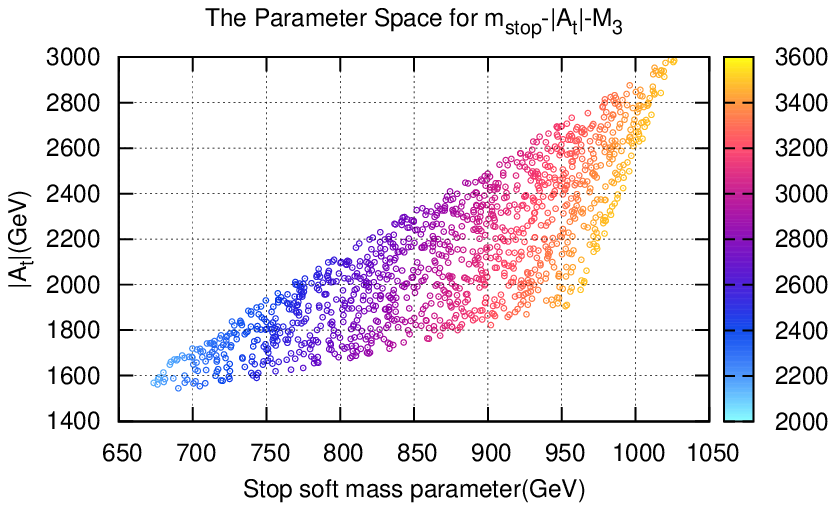},
\includegraphics[width=2.7in]{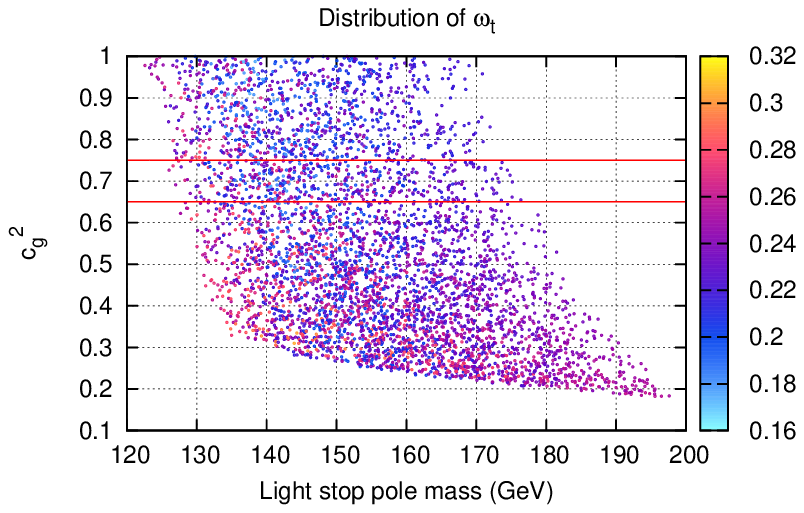}
\end{center}
\caption{\label{fig2}  Top figure: The plot of the parameter
space for $m_{\wt t_L}$-$|A_t|$-$M_3$,
where we have required $0.7<c_g^2<0.8$ and 123~GeV$<m_h<$127~GeV;
Bottom figure: The plot of distribution of $w_t$.}
\end{figure}

\noindent {\bf{Numerical search in the (N)MSSM:}} For a more precisely quantitative study,
we turn to the NMSSMTools 3.2.0~\cite{NMSSMTools}. Firstly we get the MSSM limit from the NMSSM,
e.g., setting it to the decoupling limit: $\ld=\kappa=0.01$, $\tan\beta=15$, $\mu=500$~GeV,
$A_\kappa=800$~GeV, $M_A=1.5$~TeV. One may vary them freely, but it will not bring much difference to
the configuration of the three-parameter-space $m_{\wt t_L}$-$|A_t|$-$M_3$, projected on the
$m_{\wt t_L}$-$|A_t|$ plane. From Fig.~\ref{fig2}, it is seen that the right edge of the triangle
can extend when we increase $M_3$, etc. Accordingly, the right edge of the $\omega_t$ distribution
slowly moves towards the right, where $w_t$ becomes smaller. In summary, a multi-TeV gluino,
$|A_t|$ and quite small $\omega_t$ are the main features of the parameter space in the MSSM.

Now we turn our attention to the limit which manifests the NMSSM specific effects
enhancing $m_h$~\cite{Kang:2012tn,Higgs:NMSSM}. It is characterized by $\ld\sim1$,
$\tan\beta\sim1$ and/or a small $\mu$. Especially, a small $\mu$, which is
consistent with naturalness, leads to not only the Higgs doublet-singlet (HS) mixing
pushing upward $m_h$~\cite{Kang:2012tn} but also the di-photon excess by suppressing
Br$(h\ra \bar bb)$~\cite{Kang:2012tn,dipho}. But in view of the BGF, the stop-loop
instead of the mixing effect should account for the excess. Therefore, we require
a small HS mixing angle such that $h\bar ff$ and $hVV$ are not modified much, as
forces us to accept some degree of tuning, from either a relatively large $\mu$ or a mild
cancellation in the Higgs mass matrix.

We restrict the discussion to a smaller $\mu$ and explore how natural the NMSSM
can be after requiring the BGF. Roughly speaking, this amounts to find the gluino 
lower bound~\cite{Blum:2012ii,Kang:2012tn}, which has been shown to be large.
$|A_t|$ also should be sufficiently small~\cite{Blum:2012ii,Kang:2012tn,Wymant:2012zp}. 
We give a benchmark point here. The inputs at the SUSY scale are
$M_1=110~{\rm GeV}$, $ M_2=740~{\rm GeV}$,
$M_3=1500~{\rm GeV}$, $A_t=-1530~{\rm GeV}$, and $m_{\wt t_L}=588~{\rm GeV}$.
Also,  $\tan\beta$, $\mu$ and the NMSSM specified parameters are
$\tan\beta=2.17$, $\ld=0.666$, $\kappa=0.253$,
$A_\ld=510~{\rm GeV}$, $A_\kappa=-248~{\rm GeV}$, and $\mu=291~{\rm GeV}$.
The resulting particle spectra contains the SM-like Higgs, light stop
and lightest neutralino with the dominant bino component. Their pole masses
are given by
$m_h=124.8~{\rm GeV}$, $M_{\chi_1^0}=100.0~{\rm GeV}$, and $m_{\wt t_1}=130.3~{\rm GeV}$.
In particular, $c_g=-0.84$ is generated.


We would like to make some comments. (A) The stop-chargino loop generates
flavor transition $b\ra s\gamma$. However, its rate is safely small if we take
$\tan\beta\lesssim20$ and $\mu$ a few hundred GeV~\cite{Bertolini:1990if}.
(B) Due to the large $A_t-$term, the split stop-sbottom may render $\Delta\rho$
orders larger than the experimentally allowed bound $\lesssim10^{-3}$.
Small $m_{\wt t_L}$ is favored to lower the size of $\Delta\rho$, and some
cancellation from $m_{\wt b_R}$ is needed. (C) We can relax the lower bound on
$m_{\wt t_1}$ slightly, e.g., to 125~GeV, by means of introducing $R-$parity violation
operators $U^cD^cD^c$. It makes the very light stop missed at the LHC
due to the absence of missing transverse energy.
In this case, we do not need the heavy gluino and the
BGF can be naturally realized in the NMSSM.

\noindent {\bf{A light stop escapes from the LHC:}} In both models, if $M_3<4$~TeV,
we predict $m_{\wt t_1}\lesssim m_t$, see Fig.~\ref{fig2}. As long as we
impose the minimal fine-tuning condition, it is found that the NMSSM allows
$M_3\lesssim1.5$~TeV and gives $m_{\wt t_1}\simeq130$~GeV which will be checked safe.

The LHC search for the stop is sensitive to its decay topology and the particle
spectrum. As a case in point, we consider the benchmark point in the NMSSM. It
contains the relevant particles, {\it i.e.}, the light stop with mass 130~GeV and
the 100~GeV bino-like neutralino LSP $\chi_1^0$. Due to the kinematics under
consideration, $\wt t_1$ dominantly (nearly 100$\%$) decays to $c$ and $\chi_1^0$
via the loop-induced process.
The process at the LHC is
     \begin{align}\label{}
gg\ra\wt t_1\wt t_1^*\ra c+\bar c+2\chi_1^0.
  \end{align}
In this case, the monojet and jets+$E_T^{\rm miss}$ signatures give the strongest
bounds~\cite{He:2011tp}. Actually, the monojet bound is effective only when
$0<m_{\wt t_1}-m_{\chi_1^0}<0.2m_{\chi_1^0}$. Otherwise, a putative case in
this work, it is ineffective and need the inspection to
jets$+E_T^{\rm miss}$, discussed in the following.

At first, we use MadGraph5 v1.4.7~\cite{Alwall:2011uj} to generate the stop
pair, accompanied by ($\leq3$) additional initial state radiative (ISR) jets.
In calculating this cross section, we take the modified MLM matching with
$x{\rm qcut}=30$~GeV and shower-kt scheme with QCUT=50~GeV. Then after matching
the LO production cross sections of $\wt t_1\wt t_1^*+nj$ at the 7~TeV LHC,
are displayed in the following:
  {\small   \begin{align}\label{}
\begin{tabular}{c | c | c | c | c}
   Channel     &   $\tilde{t} \bar{\tilde{t}}$   &   $\tilde{t} \bar{\tilde{t}}$
   + j & $\tilde{t} \bar{\tilde{t}}$ + jj  &  $\tilde{t} \bar{\tilde{t}}$ + $(\geqslant3j$) \\\hline
    Cross section (pb)   & 42.08  & 15.73  & 3.84 & 0.87
\end{tabular}.
  \end{align}}Moreover, the k-factor calculated by the Prospino2~\cite{Beenakker:1996ed} is about 1.5
for $m_{\wt t_1}\sim130$~GeV. After these, we use Pythia to do the particle decay, parton showering
and hadronization~\cite{Sjostrand:2006za}. PGS is chosen to simulate the detector effect.
Finally, employing the cuts adopted in Ref~\cite{Atlas}, we find that only the events with
two or more ISR jets survive. The most sensitive signal regions and the corresponding event
numbers (normalized to 4.7/fb) are given by:
{\small\begin{align}\label{}
\begin{tabular}{  c | c | c | c | c  }
Signal Region  & {SRC-med} & SRC-loo  & SRE-med & SRE-loo \\\hline
Number of Event & 2.03(18) & 15.10(58) & 1.02(12) & 4.07(84)
\end{tabular},\nonumber
  \end{align}}where the numbers in the brackets are the observed upper limits on
the event numbers beyond the SM. Clearly, the present ATLAS
jets$+E_T^{\rm miss}$ search can not exclude our light stop scenario. We are awaiting
the ongoing LHC to probe such scenarios.

\noindent {\bf{Conclusions and discussions}:} Stops are potential to accommodate
the BGF. At first glances, due to the relations among the competing $c_g$,
$m_{\wt t_1}$, and $m_h$, there is merely a sporting chance to realize the BGF in
the MSSM and NMSSM. Fortunately, the gluino radiative correction on the light stop mass
can make its pole mass much lighter than the running mass, and then the BGF can be
realized.
If we reasonably let $M_3$ and $|A_t|$ below 4~TeV,
a light stop typically below the top mass is predicted. In particular, in the NMSSM
we explore the less fine-tuned region to the most extent, and find that the
model allows $M_3\simeq1.5$~TeV as well as $m_{\wt t_1}\approx130$~GeV. Such light stop
is not excluded by the ATLAS latest data since the decay ${\wt t_1}\ra c+\chi_1^0$ with
$\chi_i^0$ about 100~GeV. We encourage the LHC experiments to tailor a search
for the light stop inspired by the Higgs BGF. Moreover,
besides the light stop, the lightest neutralino, which may be an successful dark matter
candidate, also deserves further attention. In our scenario, its mass falls into
the region $m_{\wt t_1}>M_{\chi_1^0}>0.83m_{\wt t_1}$, {\it i.e.}, between 100~GeV
and the $m_t$.
We leave it for a future study. Finally, after the completion of this work we are aware of
~\cite{Reece:2012gi}, which gives a stringent constraint on the BGF inspired stop sector 
from the vacuum stabilities. But their constraint is based on a rather simplified analysis 
and only valid at the tree-level.

\noindent {\bf{Acknowledgements}:}
   This work was supported by the
National Natural Science Foundation of China under grant Nos.
10821504, 11075194, and 11135003,
and by the DOE grant
DE-FG03-95-Er-40917.

\end{document}